\documentclass{proceedingsM}

\usepackage[hidelinks]{hyperref}  
\usepackage{float} 
\usepackage{enumerate} 
\usepackage[numbers,comma,sort&compress]{natbib}

\title{Temperature changes accompanying signal propagation in axons} %

\author{Kert Tamm\corauthref{Corresponding author}, J\"uri Engelbrecht, Tanel Peets}

\address{Department of Cybernetics, School of Science, Tallinn University of Technology, Akadeemia tee 21, Tallinn 12618, Estonia}
\email{kert@ioc.ee, je@ioc.ee, tanelp@ioc.ee}

\abstract{In this paper mathematical models are formulated in order to simulate heat production and corresponding temperature changes which accompany the propagation of an axon potential. Based on earlier experimental results, several models are proposed. Together with the earlier system of coupled differential equations derived by the authors for describing the electrical and mechanical components of signalling in nerve fibres, the novel results permit to cast the whole process of signalling into one system. The emphasis  is on the mathematical description of coupling forces. The numerical results are qualitatively similar to experiments. }


\begin{document}

\maketitle


\section{INTRODUCTION}
The generation of heat accompanying the propagation of an action potential has been measured in many experimental studies by Abbott et al \cite{Abbott1958}, by Downing et al \cite{Downing1926}, by Howarth et al \cite{Howarth1968}, by Ritchie et al \cite{Ritchie1985}, by Tasaki et al \cite{Tasaki1988,Tasaki1992} etc. However, there is no generally accepted  mechanism for heat production and temperature changes because the measurements differ in a large scale. The generally accepted energy of a membrane capacitor is related to the square of voltage \cite{Ritchie1985}. Tasaki and Byrne \cite{Tasaki1992} have shown that either the generated temperature or its rate may be related to voltage depending on physical properties. It is of great interest to formulate the mathematical basis for calculating the heat production or temperature changes given these experimental assumptions. In our earlier studies \cite{Engelbrecht2018c,Engelbrecht2018,EngelbrechtTammPeets2014,Engelbrecht2018d,Peets2015} a system of differential equations is derived in order to model the action potential (AP) and accompanying mechanical effects -- pressure wave (PW) in axoplasm, longitudinal (LW) and transverse (TW) waves in the surrounding biomembrane. This system will here be enlarged by including the temperature ($\Theta$) effects. The model equations are presented with the focus on possible models of heat and temperature production in Section \ref{MEQ} Further on, the numerical examples are given in dimensionless setting in Section \ref{NUMRES} The final remarks are presented in Section \ref{FINMARK}

\section{MODEL EQUATIONS}
\label{MEQ}
The coupled system of model equations has been previously proposed by the authors for describing the nerve pulse (action potential (AP)) propagation including the accompanying mechanical effects (pressure wave (PW) and longitudinal density change in the lipide bi-layer (LW)) \cite{Engelbrecht2018c,Engelbrecht2018,EngelbrechtTammPeets2014,Engelbrecht2018d,Peets2015}. 
In this paper, the temperature effects which accompany the propagation of an AP in the axoplasm, are also described. The dimensionless 1D model used so far \cite{Engelbrecht2018c,Engelbrecht2018,EngelbrechtTammPeets2014,Engelbrecht2018d,Peets2015} is the following:
\begin{equation}
\label{FHNeq}
\begin{split}
& Z_{T} = D Z_{XX} + Z \left( Z - \left[ a_1 + b_1 \right] - Z^2 + \left[ a_1 + b_1 \right] Z \right) - J, \\
& J_{T} = \varepsilon \left( \left[ a_2 + b_2 \right] Z - J \right),
\end{split}
\end{equation}
\begin{equation}
\label{iHJeq}
U_{TT} =  c^2 U_{XX} + N U U_{XX} + M U^2 U_{XX} + N U_{X}^{2} + 2 M U U_{X}^{2} - H_1 U_{XXXX} + H_2 U_{XXTT} + F_1(Z,J,P),
\end{equation}
\begin{equation}
\label{Peq}
P_{TT} = c_{f}^{2} P_{XX}  - \mu P_T + F_2(Z,J,U),
\end{equation}
where Eq.~\eqref{FHNeq} is the FitzHugh--Nagumo (FHN) equation \cite{Nagumo1962}, Eq.~\eqref{iHJeq} is the improved Heimburg--Jackson (iHJ) equation \cite{Heimburg2005,HJ2007,EngelbrechtTammPeets2014}, Eq.~\eqref{Peq} is the classical wave equation with the added viscous dampening term. 
In Eq.~\eqref{FHNeq} $Z$ is potential, $J$ is the abstracted ion current (like characteristic to the FHN equation), $a_i, b_i$ are ``electrical" and ``mechanical" activation coefficients, $D, \varepsilon$ are coefficients. In Eq.~\eqref{iHJeq} $U$ is longitudinal density change, $c$ is sound velocity of unperturbed state in the lipid bi-layer, $N, M$ are nonlinear coefficients, $H_1, H_2$ are dispersion coefficients. In Eq.~\eqref{Peq} $P$ is pressure, $c_{f}$ is sound velocity in axoplasm, $\mu$ is dampening coefficient.  Here and further index $X$ denotes spatial partial derivative and index  $T$ denotes partial derivative with respect to time.

Coupling terms have been described by polynomials
\begin{equation}
\label{Forces}
F_1 = \gamma_1  \frac{P_T}{1+U} + \gamma_2 \frac{J_T}{1+U} - \gamma_3 \frac{Z_T}{1+U};
\quad
F_2 = \eta_1 Z_X + \eta_2 J_T + \eta_3 Z_T,
\end{equation}
where $\gamma_i$ are coupling coefficients for mechanical wave and  $\eta_i$ are coupling coefficients for the pressure wave. Time derivatives act across membrane and spatial derivatives act along the axon.

What has not been considered in the previous publications on mathematical modelling is the possible temperature change accompanying the AP. As far as system \eqref{FHNeq} -- \eqref{Peq} is a system of partial differential equations (PDE), we look for an additional equation capable to model the temperature changes. The crucial question is how temperature effects are related to other propagating effects and as before we look for corresponding coupling forces which in this case should be related to the heat production.
It is worth also keeping in mind that temperature change normally is not considered like a wave but is considered to be like a diffusive process. 
There are several ideas proposed in earlier studies on thermal changes caused by propagating APs.
It has been stated that the energy of the membrane capacitor $E_c$ is \cite{Ritchie1985}
\begin{equation}
\label{Eceq}
E_c = \frac{1}{2} C_m Z^2,
\end{equation}
where $C_m$ is a capacitance and $Z$ is the amplitude of the AP. Noting that heat energy $Q \approx E_c$, it can be deduced that $Q$ should be proportional to $Z^2$. The standard Fourier law (the thermal conductivity equation) in its simplest 1D form is
\begin{equation}
\label{FourierL}
Q = -k \Theta_X,
\end{equation}
where $Q$ is the heat energy, $k$ is the thermal conductivity and $\Theta$ is the temperature. Note that heat energy is proportional to the negative temperature gradient. Combining \eqref{Eceq} and \eqref{FourierL}, we obtain 
\begin{equation}
\label{ThetaEq}
\Theta \propto  \frac{C_m}{2 k} \int Z^2 dX.
\end{equation}
Further we follow experimental results by Abbot et al \cite{Abbott1958} which demonstrate that the heat increase at the surface of the fibre is positive. 
It has also been argued that either $Z \propto \Theta$ or $Z \propto \Theta_T$ depending on physical properties \cite{Tasaki1992}. The potential $Z$ of the AP or its square $Z^2$ might be not the only sources. 
For example, in \cite{Heimburg2008,Schneider2018} it is argued in favour of the idea that experimentally observed temperature changes might be the result of the propagating mechanical wave in the lipid bi-layer. As there seems to be no clear consensus which quantities associated with the nerve pulse propagation are the sources of the thermal energy, we will consider here all the possibilities for the coupling of the waves in the ensemble to the additional model equation for the temperature. 

The idea is to cast these ideas into a mathematical form.
As far as temperature is a function of space and time
we opt to use the classical heat equation  which is a parabolic PDE describing the distribution of heat (or variation in temperature) in a given region over time. 
It is straightforward \cite{CarslawJaeger1959} to derive the heat equation in terms of temperature from the Fourier's law by considering the conservation of energy 
\begin{equation}
\label{HHeq}
\Theta_{T} = \alpha \Theta_{XX},
\end{equation}
where $\alpha$ is the thermal diffusivity. In our case, a possible source term must be added to Eq.~\eqref{HHeq}
\begin{equation}
\label{Heq}
\Theta_{T} = \alpha \Theta_{XX} + F_3(Z,J,U).
\end{equation}
The source term $F_3$ permits to account for different assumptions proposed to describe the temperature generation and consumption.
Further on, a number of different functions $F_3$ are  used for this purpose: 
(i) $F_3 = \tau_1 Z$ and  $F_3= \tau_2 Z^{2}$ (see Fig.~\ref{Fig2});
(ii) $F_3 = \tau_3 J$ and  $F_3= \tau_4 J^{2}$ (see Fig.~\ref{Fig3});
(iii) $F_3 = \tau_5 U$ and  $F_3= \tau_6 U^{2}$ (see Fig.~\ref{Fig4});
(iv) $F_3 = \tau_7 Z_T + \tau_8 J_T$ and $F_3 = \tau_9 J_T + \tau_{10} U_X,$ (see Fig.~\ref{Fig5})
where $\tau_i$ are the thermal coupling coefficients. 

\section{NUMERICAL EXAMPLES AND DISCUSSION}
\label{NUMRES}
The model equations are solved using the pseudospectral method (PSM) \cite{Engelbrecht2018d}. 
The added heat equation \eqref{Heq} can be directly solved as outlined in  \cite{Engelbrecht2018d} following the same procedure without further modifications. The idea of the PSM is write the PDEs in a form where all the time derivatives are on left hand side (LHS) and all the spatial derivatives on the right hand side (RHS) of the equation and then apply properties of the Fourier transform for finding the spatial derivatives reducing the PDE into an ordinary differential equation (ODE) which can be solved by standard numerical methods. For the numerical solving -- a localized initial condition is used in the middle of the spatial period for $Z$ with the rest of the processes assumed to be at rest (zero initial conditions). This initial ``spark" is taken above the threshold value in the middle of spatial period causes the AP to emerge and propagate to the left and right. All other waves are generated by the propagating AP as a result of coupling terms added to the system. Boundary conditions are taken periodic as is required for the PSM, moreover in the present paper the time integration intervals have been taken short enough to avoid interaction between counter-propagating waves at the boundaries. In the figures only left propagating waves are plotted. 

The following parameter values have been used for the numerical simulations: \vspace{-3mm}
\begin{enumerate}[(i)]
\item for the FHN equation \eqref{FHNeq} \\
$
D=1;
\varepsilon=0.018;
a_1=0.2;
a_2=0.2;
b_1=-0.05 \cdot U;
b_2=-0.05 \cdot U,
$
\vspace{-3mm}
\item for the improved Heimburg-Jackson model \eqref{iHJeq} \\
$
c^2 = 0.10;
N=-0.05;
M=0.02;
H_1=0.2;
H_2=0.99;
$
\vspace{-3mm}
\item for the  pressure \eqref{Peq} \\
$
c_{f}^{2}=0.09;
\mu=0.05;
$
\vspace{-3mm}
\item for the heat equation \eqref{Heq} \\
$
\alpha=0.05;
$ 
\vspace{-3mm}
\item and the coupling coefficients are \\
$
\gamma_1=0.008; 
\gamma_2=0.01; 
\gamma_3=3 \cdot 10^{-5}; 
\eta_1=0.005; 
\eta_2=0.01; 
\eta_3=0.003;\\ 
\tau_{1, \ldots , 6} = 5 \cdot 10^{-5};
\tau_7 = 5 \cdot 10^{-5}; 
\tau_8=1 \cdot 10^{-3};
\tau_9 = 1 \cdot 10^{-3};
\tau_{10} = 5 \cdot 10^{-5}.
$
\end{enumerate}\vspace{-3mm}
The parameters for the numerical scheme are $L=64\pi$ (the length of the spatial period), $T_f = 400$ (the end time for integration, figures in the present paper are shown at $T=400$), $n=2048$ (the number of spatial grid points), the initial amplitude for $Z$ is $A_z=1.2$ and  width of the initial sech$^2$ type pulse  is $B_o=1.0$ \cite{Engelbrecht2018d}.

\begin{figure}[h]
\includegraphics[width=0.42\textwidth]{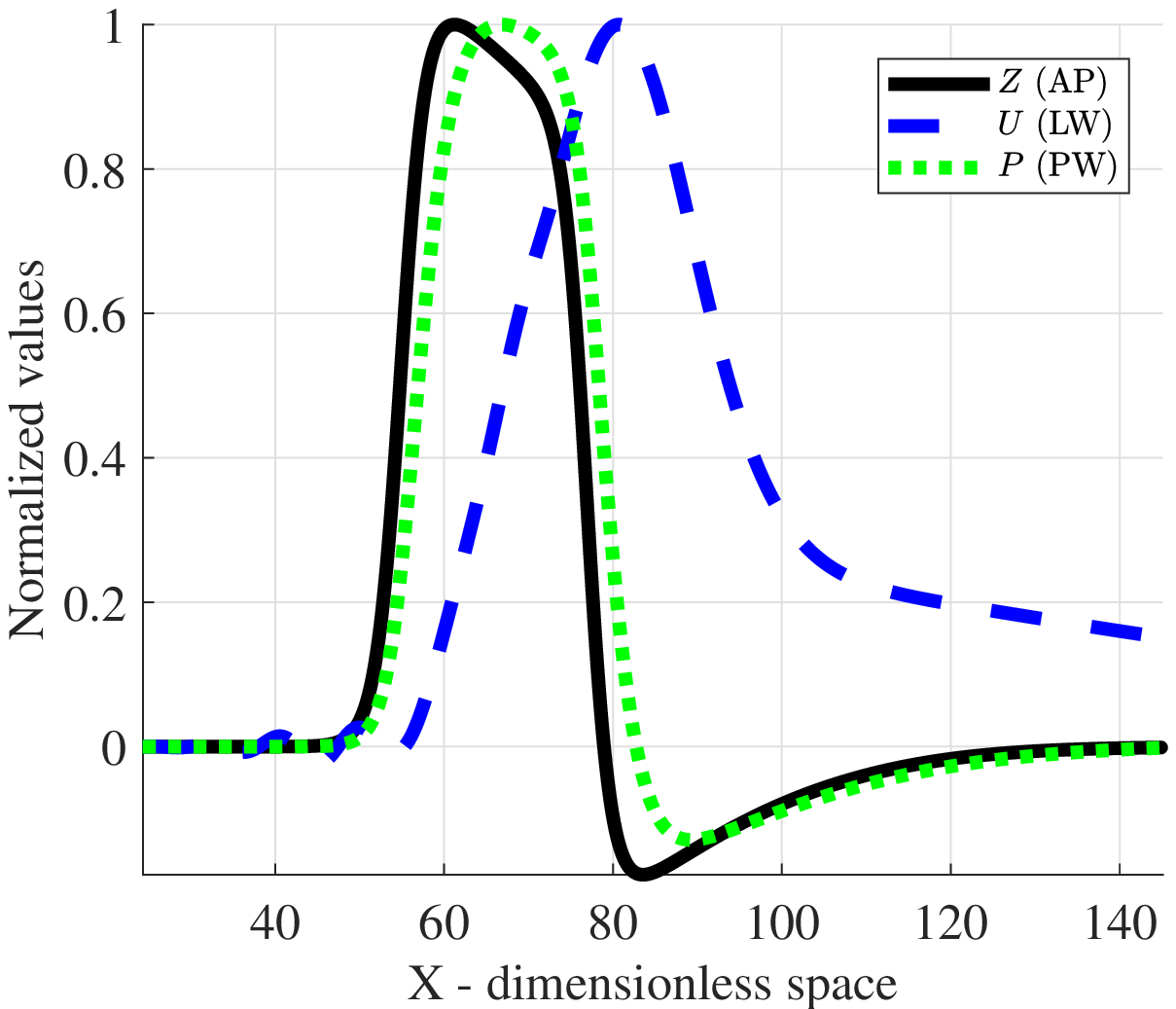}
\includegraphics[width=0.42\textwidth]{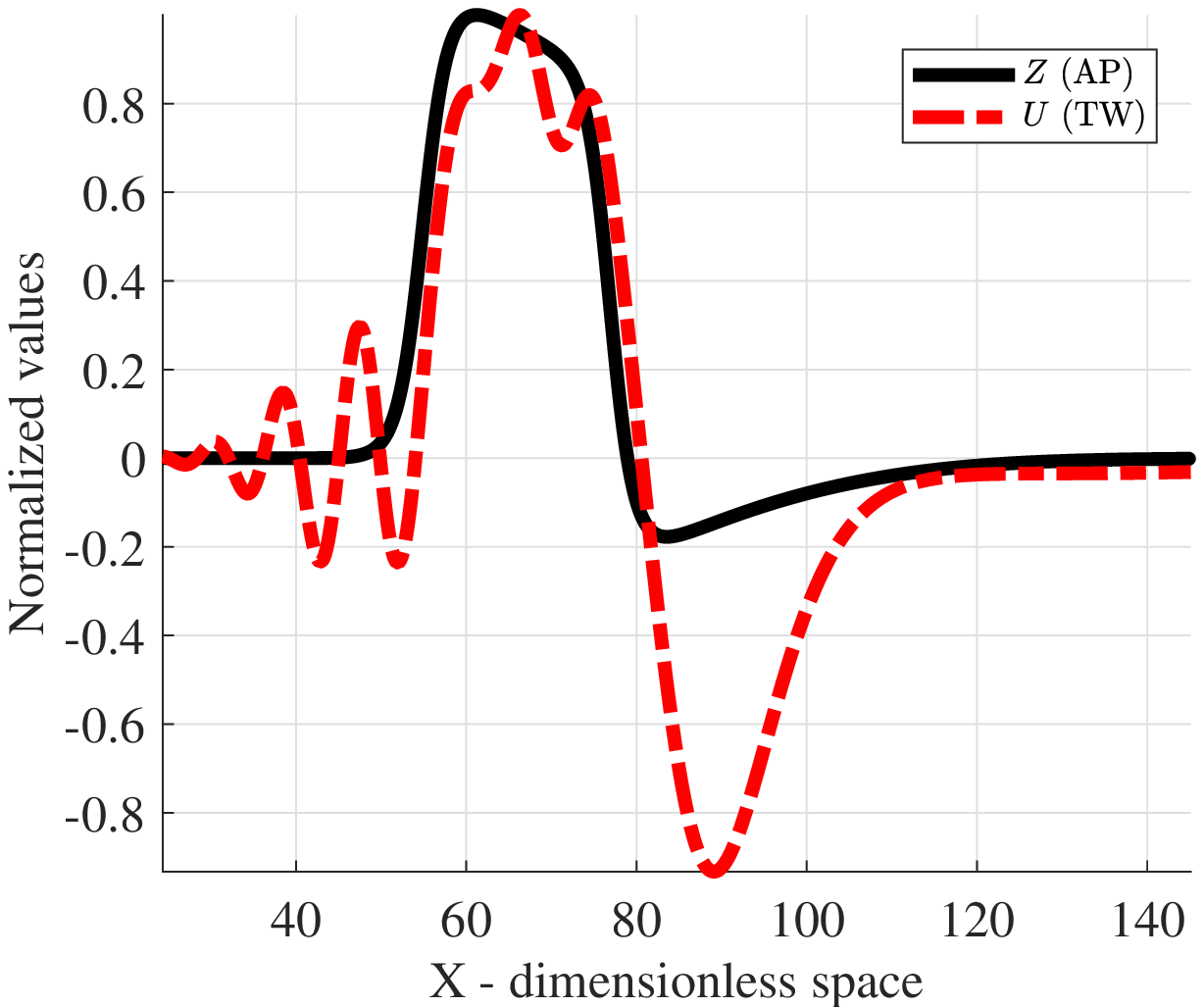}
\caption{The wave ensemble modelled quantities (left) and the TW derived from the LW (right).}
\label{Fig1}
\end{figure}

The wave ensembles are shown in Fig.~\ref{Fig1}. It should be noted that some characteristics normally observed in experiments, like the transverse displacement (TW) of the membrane is not described by a differential equation but can be derived directly from the calculated LW as a derivative \cite{EngelbrechtTammPeets2014}. Certain oscillations of the calculated TW reflect the physical properties of the biomembrane which is described by a Boussinesq-type equation \cite{EngelbrechtTammPeets2014}. 

\begin{figure}[h]
\includegraphics[width=0.42\textwidth]{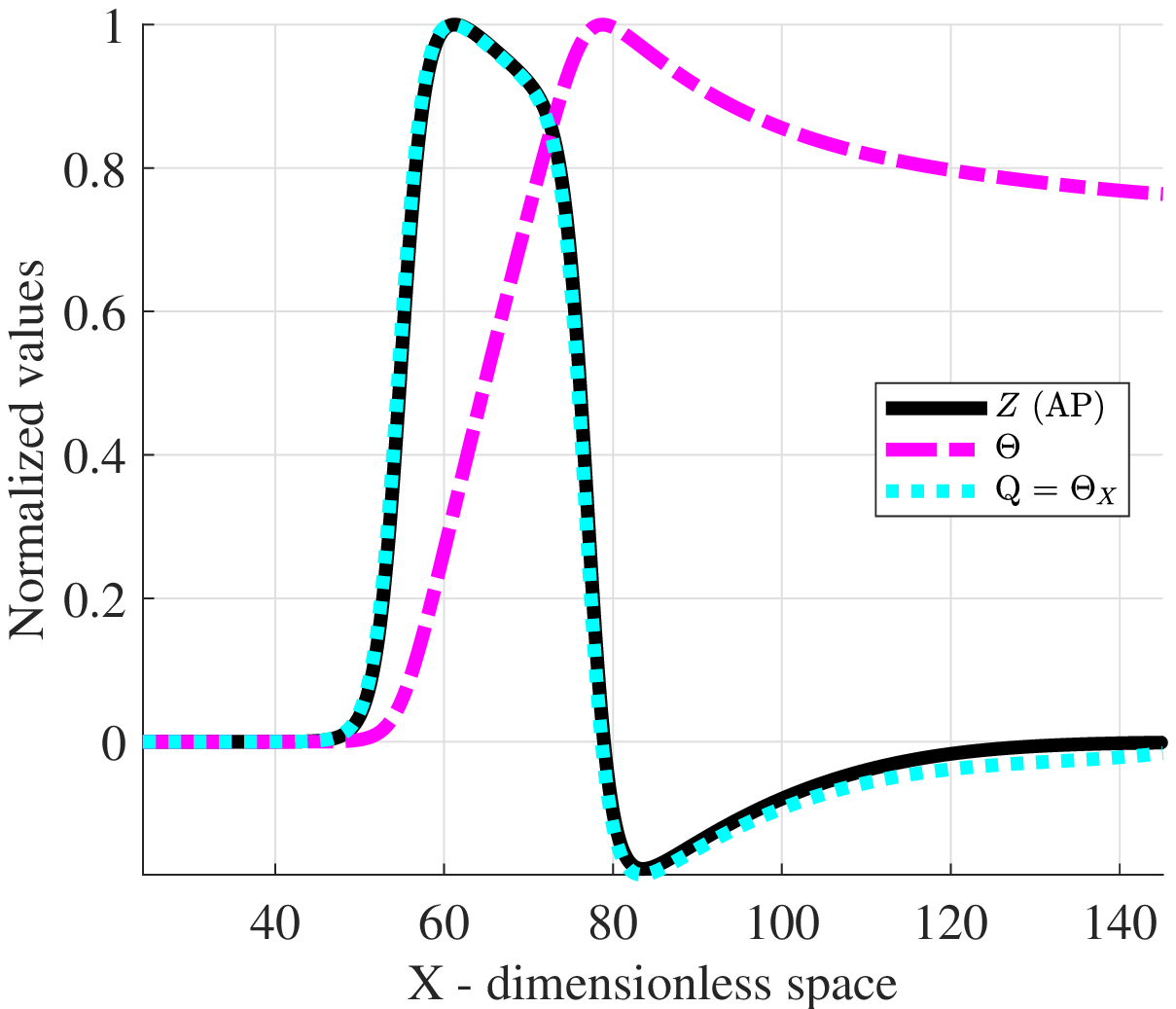}
\includegraphics[width=0.42\textwidth]{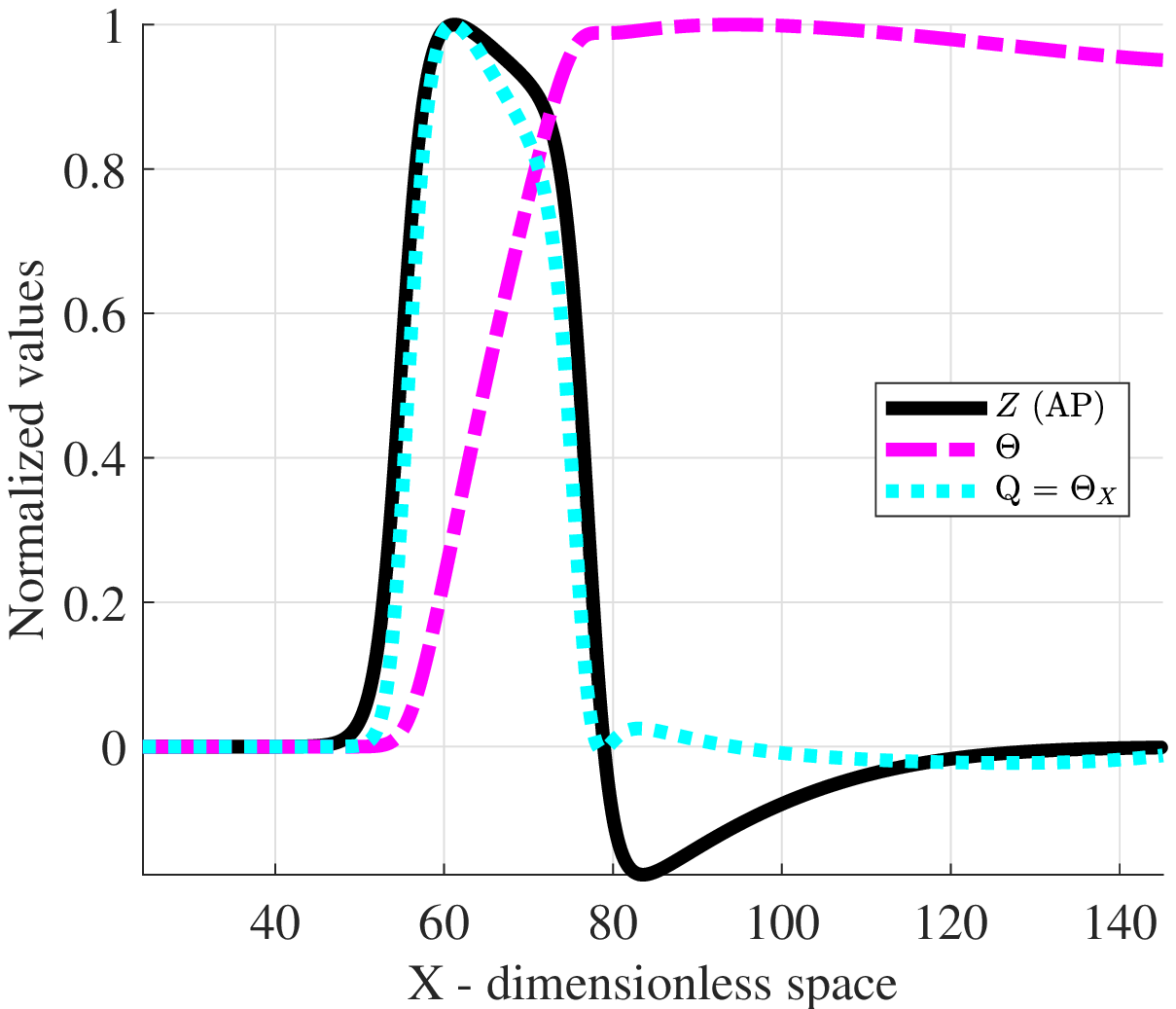}
\caption{The temperature change $\Theta$ and thermal energy change $Q = \Theta_X$ if $F_3 = \tau_1 Z$ (left) and if $F_3 = \tau_2 Z^2$(right).}
\label{Fig2}
\end{figure}

\begin{figure}[h]
\includegraphics[width=0.42\textwidth]{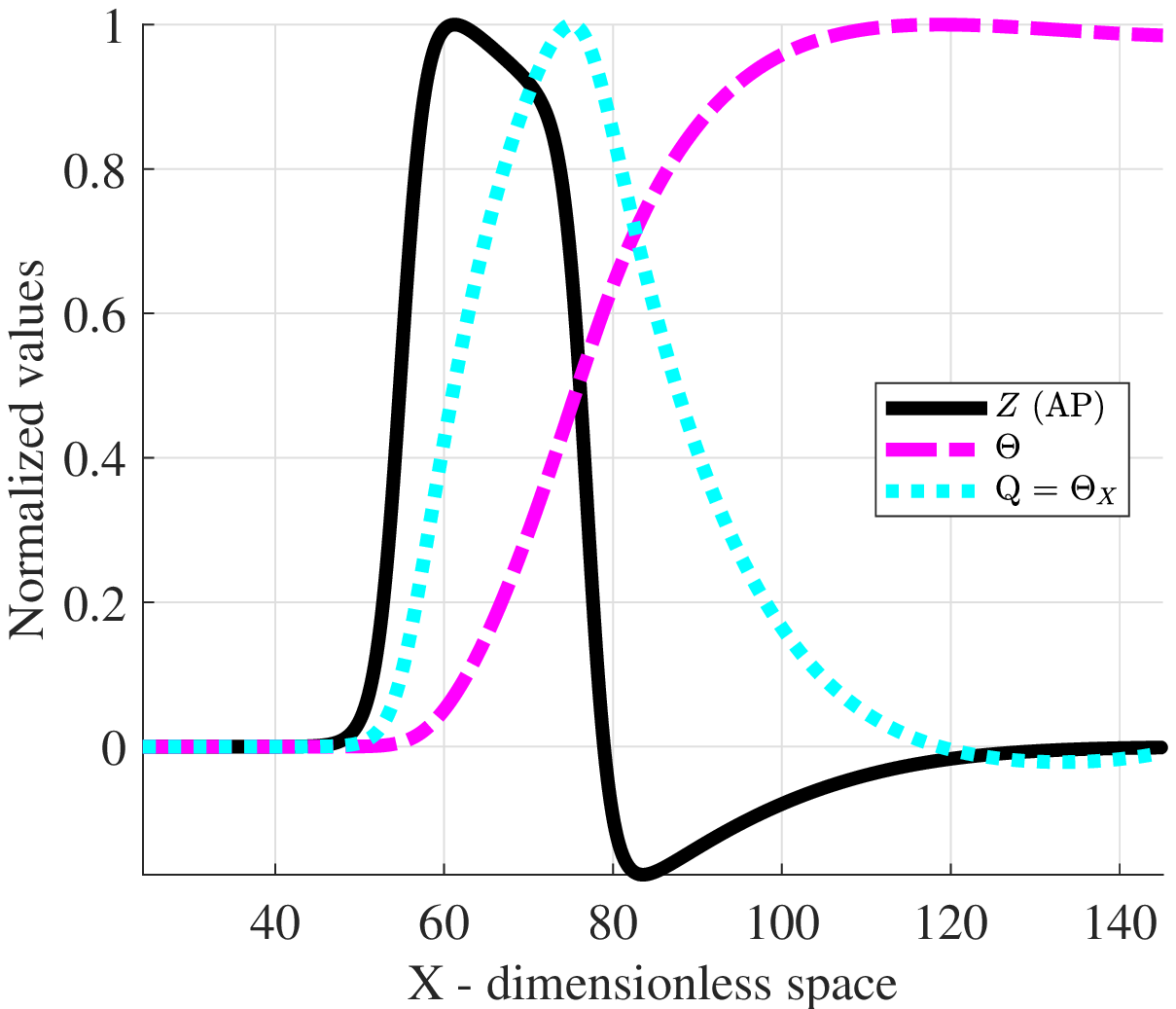}
\includegraphics[width=0.42\textwidth]{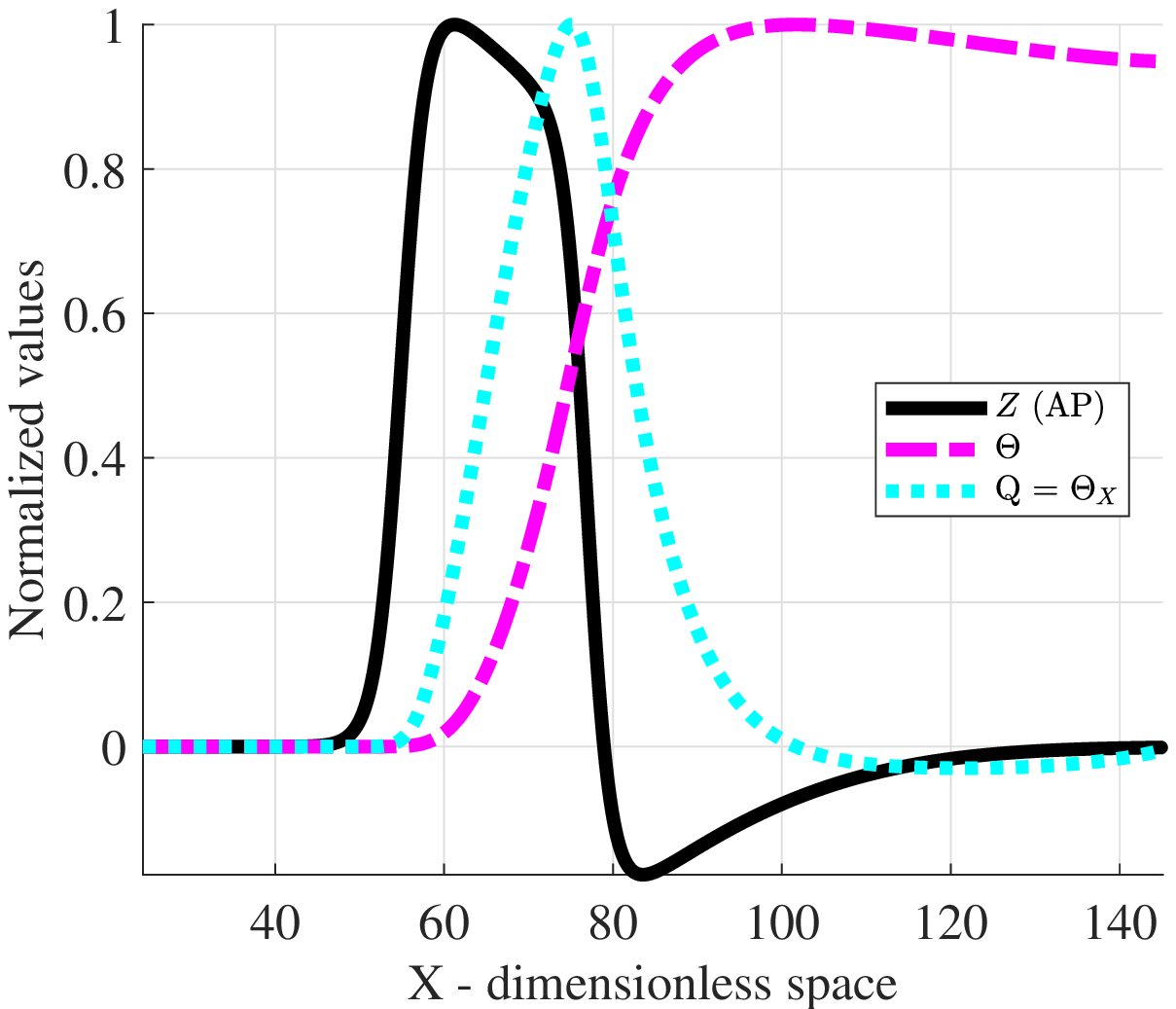}
\caption{The temperature change $\Theta$ and thermal energy change $Q = \Theta_X$ if $F_3 = \tau_3 J$ (left) and if $F_3 = \tau_4 J^2$(right).}
\label{Fig3}
\end{figure}

\begin{figure}[h]
\includegraphics[width=0.42\textwidth]{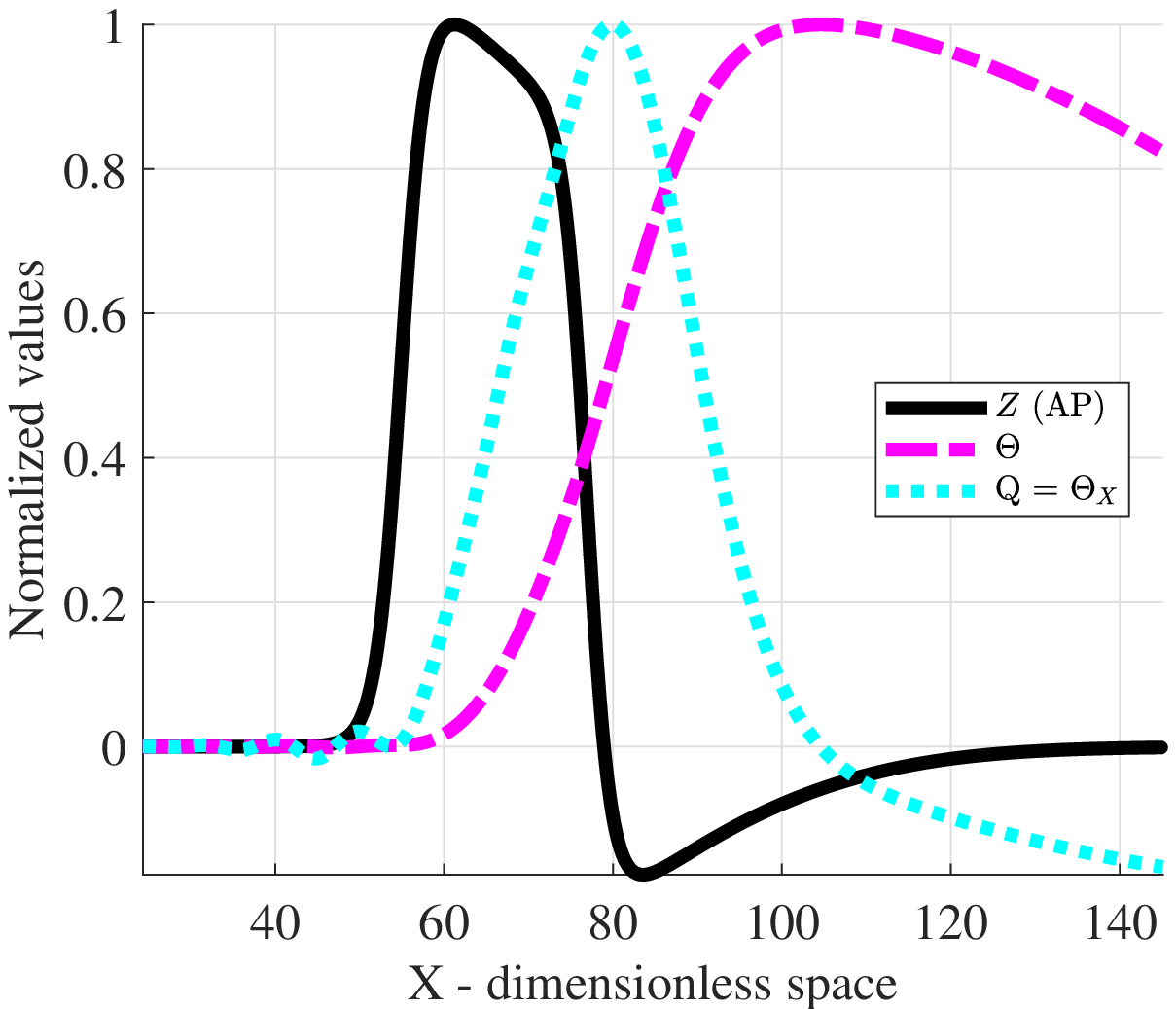}
\includegraphics[width=0.42\textwidth]{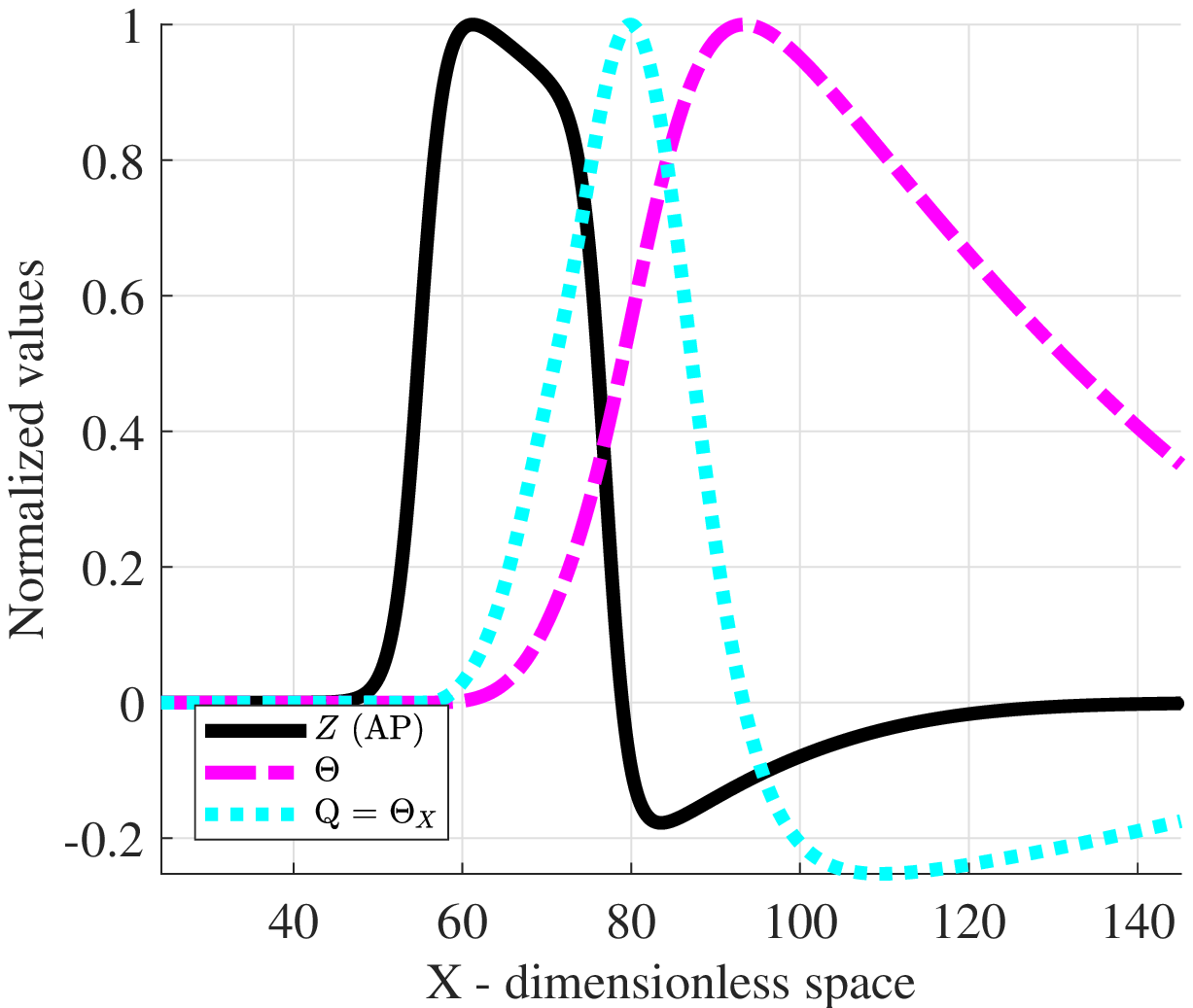}
\caption{The temperature change $\Theta$ and thermal energy change $Q = \Theta_X$ if $F_3 = \tau_5 U$ (left) and if $F_3 = \tau_6 U^2$(right).}
\label{Fig4}
\end{figure}

\begin{figure}[h]
\includegraphics[width=0.42\textwidth]{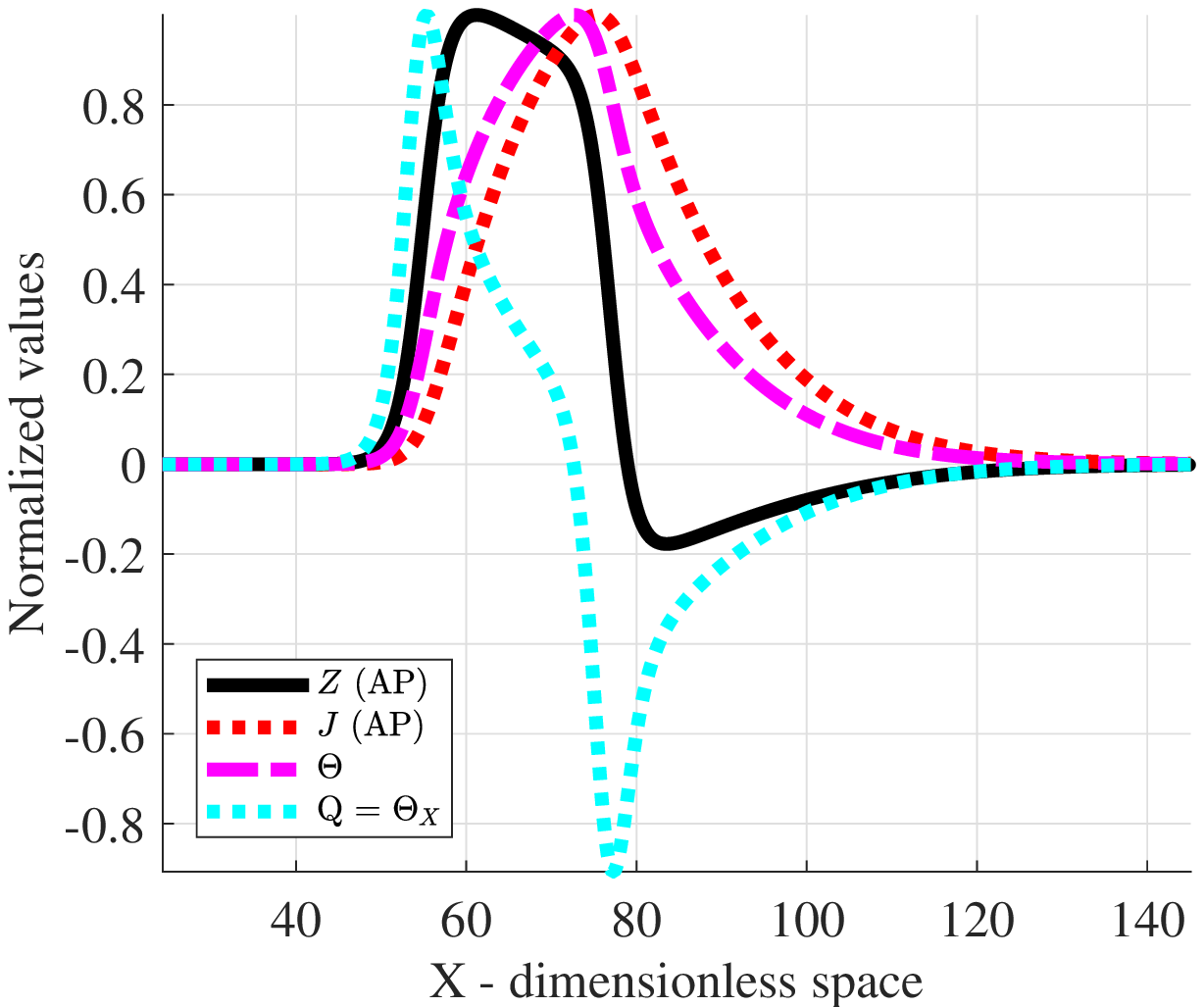}
\includegraphics[width=0.42\textwidth]{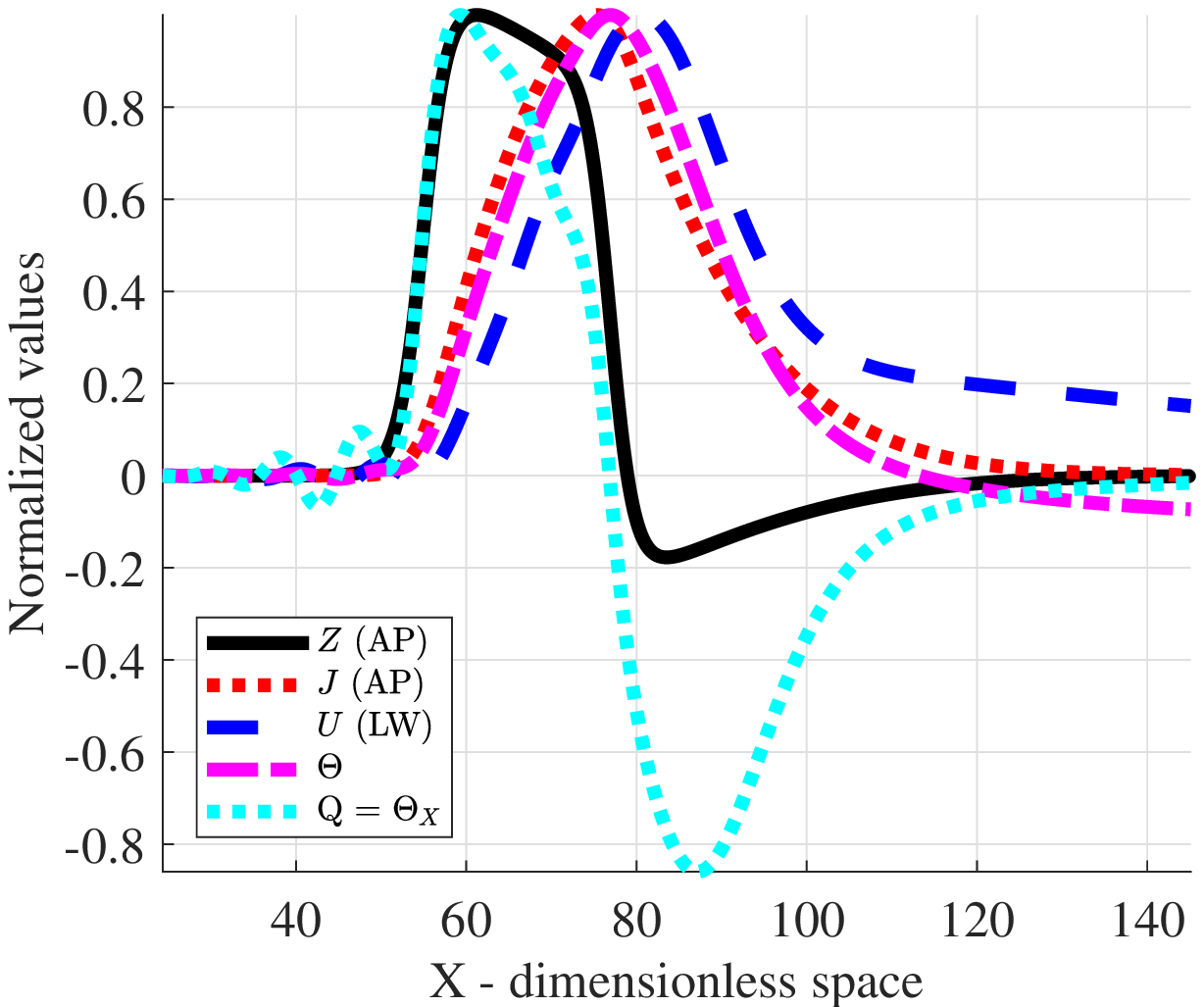}
\caption{The temperature change $\Theta$ and thermal energy change $Q = \Theta_X$ if $F_3 = \tau_7 Z_T + \tau_8 J_T$ (left) and if $F_3 = \tau_9 J_T + \tau_{10} U_X$(right).}
\label{Fig5}
\end{figure}


Few more remarks are in order. Note that in Figs \ref{Fig2} -- \ref{Fig5} thermal energy change $Q$ is plotted as $Q = \Theta_X$. 
One can observe in Fig.~\ref{Fig2} that in the case of temperature increase proportional to $Z$ the local temperature drops in the negative polarity region of the AP while if the temperature increases proportional to $Z^2$ the local temperature keeps increasing even in the negative polarity region of the AP. Observing correlation between the $Z$ (or $Z^2$) and the measured temperature increase near axon seems to be common for several experiments by Howarth et al \cite{Howarth1968}, Ritchie et al \cite{Ritchie1985} and by Tasaki et al \cite{Tasaki1992}. Experimental results published so far seem inconclusive to argue firmly in favour of either option as normally only some kind of averaged thermal energy production over some time can be measured. Another note made is that while experimentally the heat production proportional to AP has been observed, this correlation does not have to mean direct causality. Like generally in nature, the potential gradient alone is rarely the only source of the temperature change in an environment. One could argue that heat production might be instead be proportional to ion currents (Fig.~\ref{Fig3}) or the longitudinal density change (Fig.~\ref{Fig4}) which accompany the propagating AP. For the sake of completeness temperature increase as a function of $J$ and $J^2$ is presented in Fig.~\ref{Fig3} and as a function of density change $U$ and $U^2$ in Fig.~\ref{Fig4}. The idea behind including $J$ and $U$ as sources is that mechanisms where the temperature increase is caused by the current flowing through the environment or by the deformation of the solid are well established in the physics even if these are not as common sources assumed for the heat generation \cite{Heimburg2008,Schneider2018} as the correlation with the AP \cite{Howarth1968,Ritchie1985,Tasaki1992}.

In such a context it might not be a problem to observe that some heat energy ``goes away" in the negative polarity phase of the used driving signal if one assumes heat production proportional to the driving signal as the underlying endothermic mechanism might be actually independent of the AP propagation, like, some kind of endothermic chemical reaction starting \cite{Abbott1958}. The open question in the latter case is, however, the issue of time scales. Moreover, another effect present in the numerical simulation results presented must be noted -- at least in part the reason why the temperature can be lower in the areas through which the signal has already propagated as opposed to the front is that the driving signal can be actually of lower amplitude at earlier stages of the signal propagation. This is the main reason why in Fig.~\ref{Fig4} there is quite a significant drop in temperature for the case of $U^2$ (right panel) because in this case the change in driving signal amplitude is amplified in time. 
The signal shape in Fig.~\ref{Fig4} (right panel) is qualitatively similar to the signal shapes observed by Abbott, et al in \cite{Abbott1958}. 
Another note relevant specifically for the Eq.~\eqref{iHJeq} can be made -- it is a conservative equation before adding the coupling term $F_1$ facilitating energy exchange between the coupled equations. However, if the mechanical wave in the lipid bi-layer is considered as a source for thermal energy in the system for earnest this equation would need some kind of additional term which would take some energy away from the mechanical wave. For example, the simplest possibility might be using a Voigt--type model which means appearing a term  $U_{XXX}$ in Eq.~\eqref{iHJeq}. 

Experimentally the thermal energy production and consumption has been observed in the time scales comparable to the AP propagation \cite{Tasaki1988,Tasaki1992} to few orders of magnitude longer as shown by Howarth et al \cite{Howarth1968}. Following the idea that the underlying processes responsible for the heat production and consumption might be proportional to some of the processes we are actually modelling one possibility shown in Fig.~\ref{Fig5} where time derivatives of AP and ion current J (left) or the ion current time derivative and the gradient of longitudinal density change (right) are used for the thermal energy generation and consumption. In such a case thermal energy is generated when the driving signal is growing and consumed when the driving signal is decreasing. The examples shown in Fig.~\ref{Fig5} are intended to highlight what is mathematically possible in the present framework and not as an implication of the thermal response being actually generated and even more importantly, consumed, by the used signals in the reality. However some of the experimental observations are qualitatively similar to profiles shown in Fig.~\ref{Fig5} \cite{Tasaki1992}. A minor remark on whether to use spatial or temporal derivatives and signals is in order -- in our 1D setting the signal shape is practically the same regardless if observed in space (at fixed time) or in time (at fixed spatial point). Actually, integrating Eq.~\eqref{Heq} in time with $F_3 \propto Z^2$ (Fig.~\ref{Fig2}, right panel) takes into account expression \eqref{ThetaEq}.

Finally it seems prudent to emphasize where, exactly, is this temperature change modelled on axon in the presented framework as normally the experimental measurements for the phenomenon are performed close to the axon wall outside of the axon itself. The temperature changes in Figs \ref{Fig2}, \ref{Fig3} and \ref{Fig5} (left panel) are for the axoplasm inside the axon. The temperature plotted in Fig.~\ref{Fig4} would have to be inside the lipid bi-layer forming the axon wall because that is where the density change $U$ is taken into account in our model which is used to generate these changes in Fig.~\ref{Fig4}.  Overall, it is assumed that any temperature changes measured sufficiently close to the lipid bi-layer forming the axon wall are proportional to the temperature changes in the axoplasm or in the lipid bi-layer giving us, in essence, a combined reading of all available temperature changes in close proximity of the hypothetical measuring point (which is the case for Fig.~\ref{Fig5} right panel). In the present framework a possible time delay between heat being generated or consumed and the ``measured" signal a small distance away is not taken into account.

One possible extension for the system \eqref{FHNeq} -- \eqref{Peq}, \eqref{Heq} is to include a separate PDE into the system for capturing the temperature decrease in a manner which better aligns with some of the experimental results. For example, the temperature response can happen on a noticeably larger time scale than the AP propagation, implying potentially a different underlying mechanism as described by Abbott et al \cite{Abbott1958} and  Howarth et al \cite{Howarth1968}.

\section{FINAL REMARKS}
\label{FINMARK}
An attempt is made to cast the experimental ideas \cite{Howarth1968,Ritchie1985,Tasaki1988,Tasaki1992} into the mathematical form. The experimental results differ on a large scale -- the temperature profile follows the AP \cite{Tasaki1988} and the temperature changes are on much longer scale compared to the AP \cite{Abbott1958}. In this paper it is demonstrated that despite of such scattered results, it is possible to construct a mathematical model reproducing the experimental results. 
It seems that it is possible mathematically simulate the temperature generation as measured by Tasaki \cite{Tasaki1988} for the garfish olfactory nerve as well as the heat production measured by Howarth et al \cite{Howarth1968} for mammalian nerves.
However, the essence of coupling terms needs experimental verification. The temperature changes are governed by diffusion contrary to the wave-like behaviour of a signal in a fibre. The mathematical models described above form certainly just a basic platform for further refinements of modelling. For example, the effects of various ion currents or general oxygen consumption, CO$_2$ output, the influence of carbohydrates (see \cite{Downing1926}) should also be taken into account. 
But even within the present model, results described above may enlarge the fundamental thermodynamics of nervous signals \cite{Heimburg2008}.

\section*{ACKNOWLEDGEMENTS}
This research was supported by the European Union through the European Regional Development Fund (Estonian Programme TK 124) and by the Estonian Research Council (projects IUT 33-24, PUT 434).



%
%
%




\end{document}